\documentclass{jfm}
\usepackage{graphicx}
\usepackage{epstopdf, epsfig}

\usepackage{amstext}
\usepackage{amsmath}
\usepackage{amsfonts,amssymb}
\usepackage{framed}
\usepackage{algorithm}
\usepackage{algorithmicx}
\usepackage{algcompatible}
\usepackage{subcaption}
\usepackage{lipsum}
\usepackage{bm}
\usepackage{xcolor}
\usepackage{pdfpages}
\usepackage{lineno}

\def\dashL{\bm{\mbox{--~--~--}}}
\def\dotL{\bm{\mbox{$\cdot\ \cdot\ \cdot$}}}
\def\dashdot{- $\cdot$ -}

\DeclareMathOperator{\sign}{sgn}

\shorttitle{Numerical investigation of turbulence of surface gravity waves}
\shortauthor{Z. Zhang and Y. Pan}

\title{Numerical investigation of turbulence of surface gravity waves}

\author{Zhou Zhang\aff{1}
  \and Yulin Pan\aff{1}\corresp{\email{yulinpan@umich.edu}} }

\affiliation{\aff{1}Department of Naval Architecture and Marine Engineering, University of Michigan,
Ann Arbor, MI 48109, USA
}

\begin{document}

\maketitle
%\linenumbers
\begin{abstract}
In this paper, we numerically study the wave turbulence of surface gravity waves in the framework of Euler equations of the free surface. The purpose is to understand the variation of the scaling of the spectra with wavenumber $k$ and energy flux $P$ at different nonlinearity levels under different forcing/free-decay conditions. For all conditions (free decay, narrow- and broadband forcing) we consider, we find that the spectral forms approach wave turbulence theory (WTT) solution $S_\eta\sim k^{-5/2}$ and $S_\eta\sim P^{1/3}$ at high nonlinearity levels. With the decrease of nonlinearity level, the spectra for all cases become steeper, with the narrow-band forcing case exhibiting the most rapid deviation from WTT. To interpret these spectral variations, we further investigate two hypothetical and disputable mechanisms about bound waves and finite-size effect. Through a tri-coherence analysis, we find that the finite-size effect is present in all cases, which is responsible for the overall steepening of the spectra and reduced capacity of energy flux at lower nonlinearity levels. The fraction of bound waves in the domain generally decreases with the decrease of nonlinearity level, except for the narrow-band case, which exhibits a transition at some critical nonlinearity level below which a rapid increase is observed. This increase serves as the main reason for the fastest deviation from WTT with the decrease of nonlinearity in the narrow-band forcing case.

\end{abstract}

\begin{keywords}
Authors should not enter keywords on the manuscript, as these must be chosen by the author during the online submission process and will then be added during the typesetting process (see http://journals.cambridge.org/data/\linebreak[3]relatedlink/jfm-\linebreak[3]keywords.pdf for the full list)
\end{keywords}

\section{Introduction}\label{sec:intro}
The normal state of the ocean surface is characterized by a large number of waves at difference scales subject to nonlinear interactions in the presence of wind forcing and viscous dissipation. In such a state referred to as wave turbulence, a continuous surface elevation spectrum is usually developed with a power-law form in the inertial range, as a result of the energy cascade through scales. In the framework of weak turbulence theory (WTT), the wave spectra can be analytically computed based on the assumptions of weak nonlinearity, infinite domain and phase stochasticity, leading to the so-called Kolmogorov-Zakharov (KZ) spectra \citep{zakharov1967energy}. For surface gravity waves, the omnidirectional KZ wavenumber spectra has the form
\begin{equation}
    S_\eta(k)\propto P^{1/3}k^{-5/2},
    \label{WTT}
\end{equation}
where $P$ is the energy flux to small scales and $k$ is the wavenumber.

Since the assumptions in WTT are difficult to satisfy in finite facilities, experimental attempts to verify WTT often show deviations from \eqref{WTT}. In terms of the scaling $S_\eta(k)\sim k^\alpha$ (or its frequency counterpart), different values of $\alpha$ are observed in experiments under different conditions (e.g., forcing with different amplitudes and bandwidths), and these findings are sometimes in disagreement with one another. For example, \cite{falcon2007observation,nazarenko_lukaschuk_mclelland_denissenko_2010,deike2015role,denissenko2007gravity} show that the spectral slope $\alpha$ depends on the forcing condition and approaches \eqref{WTT} at high (or certain) forcing amplitudes. In contrast, insensitivity of $\alpha$ to forcing amplitude is reported in \cite{issenmann2013gravity,aubourg2016investigation,herbert2010observation}, but with spectral slopes steeper than \eqref{WTT}. In addition, \cite{cobelli2011different} shows that the observed spectra also depend strongly on the forcing bandwidth. The situation for the scaling between $S_\eta(k)$ and $P$ is more elusive, with \cite{falcon2007observation,issenmann2013gravity} reporting a scaling $S_\eta\sim P$ in disagreement with \eqref{WTT}. However, their measurement of $P$ is based on the mean power injected by the wave maker, which may lead to inconsistency with the concept of energy flux due to the broad-scale dissipation \citep{deike2014energy,pan2015decaying}. Using the same measurement of $P$, \cite{cobelli2011different} further suggests that $P$ and $P^{1/3}$ scalings can be realized for respectively broadband and narrow-band forcings.

The inconsistencies in experiments (with WTT and with one another) are usually attributed to factors including finite-size effect, bound waves and coherent structures. First, finte-size effect \cite[e.g.][]{pushkarev2000turbulence,lvov2006discreteness,nazarenko2006sandpile} occurs at low nonlinerity level when the nonlinear broadening is not sufficient to overcome the discreteness of $k$ caused by the finite size of the facility. This is in contrast to the continous $k$ configuration in deriving \eqref{WTT}. Second, bound waves can be considered as wave components not satisfying the dispersion relation, generated from harmonics (i.e., non-resonant interactions) or distortion of the carrier wave \citep{herbert2010observation,plant1999bound,plant2004bound}. It is found in \cite{michel2018self,campagne2019identifying} that bound waves are dominant at high frequencies and are likely responsible for the deviation of the measurements from WTT (and its dependence on forcing amplitudes). Third, coherent structures (such as rogue waves and wave breaking) occurring at high nonlinearity levels are not described by WTT, thus may lead to spectra different from \eqref{WTT}. Candidate theories to model such spectra include the Phillips spectra \citep{phillips1958equilibrium} and the Kuznetsov spectra \citep{kuznetsov2004turbulence}, which are however out of the scope of the present paper. Finally, all experiments involve other complexities such as reflection from boundaries, broad-scale dissipation and surface tension effect which inevitably affect the spectra to some extent.

In numerical simulations, we are able to have a better control of the wave field by precisely specifying the forcing/dissipation and implementation of periodic boundary conditions. This offers us a clean environment to study wave turbulence at various conditions free of the complexities that are present in experiments. Existing work include \cite{dyachenko2004weak,lvov2006discreteness} for forcing turbulence and \cite{onorato2002freely,yokoyama2004statistics} for free-decay turbulence of gravity waves in the context of Euler equations. While all these work report a scaling $S_\eta(k)\sim k^{-5/2}$ consistent with \eqref{WTT}, the simulation (in each of them) is conducted at a single nonlinearity level, and therefore is not capable of resolving/understanding the sensitivity of spectra to various conditions and their scaling with $P$. On the other hand, free-decay turbulence is studied for capillary waves in \cite{pan2014direct}, which reveals steepened spectra with the decrease of nonlinearity level. However, the finding cannot be naively applied to gravity waves, because the spectral behavior at low nonlinearity critically depends on the discrete resonant manifold \citep{hrabski2020effect} which has not been characterized for surface gravity waves.

In this work, we conduct a numerical study on the spectral properties of gravity wave turbulence at different forcing (in terms of bandwidths and amplitudes) and free-decay (with relatively broadband initial data) conditions. The purpose is to elucidate the mechanisms underlying the spectral variation under different conditions. In particular, we will focus on the hypothetical mechanisms of finite-size effect and bound waves, and will leave the study of coherent structures to our future work which directly simulates the two-phase Navier-Stokes equations (since only some of the coherent structures can be simulated in the framework of Euler equations). We also remark that while the presented findings are clear in our numerical simulations, applying them to explain the aforementioned experimental observations will necessarily require the considerations of further complexities in experiments. 

The outline and some main findings of the paper are as follows. In \S 2, we present the numerical setup of the Euler equations under forcing and free-decay conditions. In \S 3, we show the numerical results including the scaling of the wave spectra with $k$ and $P$ at different nonlinearity levels. It is found that the WTT solution is approached at high nonlinearity levels for all conditions (free-decay, narrow- and broadband forcing). The spectra deviate from WTT as the nonlinearity level decreases with the largest deviation rate observed in the narrow-band forcing case. Mechanisms leading to the variation of spectra with nonlinearity levels are discussed in terms of bound waves and finite-size effect. Through a tri-coherence analysis we find that finite-size effect is present at low nonlinearities for all cases, responsible for the overall steepening of the spectra and reduced energy flux capacity. The fraction of bound waves generally decreases with the decrease of nonlinearity, but exhibits a sharp transition and explains the rapid deviation of spectra from WTT in the narrow-band forcing case. The conclusions are provided in \S 4.

\section{Mathematical formulation}\label{sec:numer}
We consider gravity waves on a two-dimensional free surface of an incompressible, inviscid and irrotational fluid. The flow can be described by a velocity potential $\phi(\bm{x},z,t)$ satisfying the Laplace's equation. Here $\bm{x}=(x,y)$ is the horizontal coordinates, $z$ is the vertical coordinate and $t$ is time. The surface velocity potential is defined as $\phi^S(\bm{x},t)=\phi(\bm{x},z,t)|_{z=\eta}$, where $\eta(\bm{x},t)$ is the surface elevation. The evolutions of $\eta$ and $\phi^S$ satisfy the Euler equations in Zakharov form \citep{zakharov1968stability}:
\begin{equation}
    \eta_t+\nabla_{\bm{x}}\eta\cdot\nabla_{\bm{x}}\phi^S-(1+\nabla_{\bm{x}}\eta\cdot\nabla_{\bm{x}}\eta)\phi_z=0
    \label{eq:eta}
\end{equation}
\begin{equation}
    \phi_t^S+\eta+\frac{1}{2}\nabla_{\bm{x}}\phi^S\cdot\nabla_{\bm{x}}\phi^S-\frac{1}{2}(1+\nabla_{\bm{x}}\eta\cdot\nabla_{\bm{x}}\eta)\phi_z^2=0
    \label{eq:phi}
\end{equation}
where $\phi_z(\bm{x},t)=\partial\phi/\partial z|_{z=\eta}$ is the surface vertical velocity, $\nabla_{\bm{x}}=(\partial/\partial x,\partial/\partial y)$ denotes the horizontal gradient. In \eqref{eq:phi}, we assume that the mass and time units are properly chosen such that density and gravitational acceleration both take values of unity \citep{dommermuth1987high}. 

To integrate \eqref{eq:eta} and \eqref{eq:phi} in time, we use the higher-order spectral (HOS) method \citep{dommermuth1987high,west1987new}. We use a nonlinearity order $M=3$ which includes cubic nonlinear terms corresponding to four-wave interactions in spectral space, consistent with the dominant energy transferring processes for gravity waves \cite[e.g.][]{hammack1993resonant,mei2005theory}. All simulations are conducted in doubly periodic square domain of size $2\pi\times 2\pi$ corresponding to a fundamental wavenumber $k_0=1$, with spatial resolutions of $512\times512$ (that are sufficient to capture the phenomena of interest in this work). To account for dissipation at high wavenumbers, we add two artificial terms respectively on the right hand sides of \eqref{eq:eta} and \eqref{eq:phi}:
\begin{equation}
    D_\eta(k)=\gamma_k\eta,
    \label{eq:Deta}
\end{equation}
\begin{equation}
    D_{\phi^S}(k)=\gamma_k\phi^S,
    \label{eq:Dphi}
\end{equation}
with the dissipation coefficient $\gamma_k$ defined as
\begin{equation}
    \gamma_k=\gamma_0(k/k_d)^\nu,
\label{eq:dissipation}
\end{equation}
where $\gamma_0$, $k_d$ and $\nu$ are parameters characterizing the dissipation. This formulation is equivalent to the low-pass filter operation used in \cite{xiao2013study}, developed through phenomenological matching with measurement of dissipation in experiments. In current work, we use values of $\gamma_0=-50$, $k_d=115$ and $\nu=30$. We note that this formulation provides a sharp transition to dissipation range above $k\approx k_d$, which is essential for us to use dissipation rate to measure energy flux in our numerical studies (i.e., free of the broad-scale dissipation effect).    

In this work, we conduct simulations for a free-decay case, and two cases with external forcing of broad and narrow bandwidth. For the free-decay case, we use as initial condition a wavenumber spectra converted from a directional frequency spectra $S_D(\omega,\theta)=D(\theta)G(\omega)$, where $\theta$ is the directional angle with respect to the positive $x$ direction. The spreading function $D(\theta)$ characterizes the angular dependence of the spectra and is chosen to be a cosine-squared function \citep{tanaka2001verification,onorato2002freely}:
\begin{equation}
    D(\theta)=
    \left\{
    \begin{array}{lc}
    \frac{2}{\pi}\cos^2\theta,& |\theta|\leq \pi/2\\
    0,& |\theta|>\pi/2
    \end{array}.
    \right.
\end{equation}
The frequency spectra $G(\omega)$ is set to the form of a Gaussian function:
\begin{equation}
    G(\omega)=\frac{B}{\sqrt{2\pi\sigma^2}}\exp\left[-\frac{(\omega-\omega_p)^2}{2\sigma^2}\right]
\label{eq:gaussian}
\end{equation}
with $\sigma=0.4$, and $B$ a parameter characterizing the initial effective steepness $\epsilon\equiv k_pH_s/2$ (with $H_s$ the significant wave height and $k_p$ the peak wavenumber) as a measure of the nonlinearity level. We use $\omega_p=\sqrt{10}$ corresponding to $k_p=10$ in the initial wavenumber spectra. 

For the forcing cases, the initial condition is a quiescent water surface and the waves are generated by forcing with different bandwidths and amplitudes. Numerically the forcing is modeled by an artificial pressure term $Q(\theta,k,t)=H(\theta)F(k,t)$ added to the right hand side of equation \eqref{eq:phi}. The angular cut-off function $H(\theta)$ takes a value of one for $|\theta| \leq \pi/4$ and zero otherwise. (We have tested that the spreading angles in all cases do not critically affect the results presented in this paper, but using a relatively narrower angle in the forcing case favorably stabilizes the simulation.) The function $F(k,t)$ takes the form \citep[e.g.][]{dyachenko2004weak,pan2020high}
\begin{equation}
    F(k,t)=
    \left\{
    \begin{array}{lc}
    f_k\exp[-C t+i(\omega_k t+R)],& t\leq T_c \\
    f_k\exp[-C T_c+i(\omega_k t+R)],& t> T_c
    \end{array},
    \right.
\label{eq:forcing}
\end{equation}
with
\begin{equation}
    f_k=
    \left\{
    \begin{array}{lc}
    f_0\frac{(k-k_1)(k_2-k)}{(k_1-k_2)^2},& k_1\leq k \leq k_2 \\
    0,& \mathrm{otherwise}
    \end{array},
    \right.
\label{eq:forcingAmp}
\end{equation}
where $f_0$ is the parameter determining the forcing amplitude, $\omega_k$ is the angular frequency for wavenumber $k$ calculated from the dispersion relation, $k_1$ and $k_2$ are the lower and upper bounds of the forcing range, and $R$ is a random number uniformly distributed in $[0,2\pi]$ that are different for each wave mode. In the broadband and narrow-band cases, we use $[k_1,k_2]=[1,19]$ and $[k_1,k_2]=[9,11]$ respectively, both corresponding to a peak mode of $k_p=(k_1+k_2)/2=10$. As described by \eqref{eq:forcing}, the forcing level decays exponentially in time with rate $C$ before $t=T_c$, and then remains constant. This provides a fast convergence to stationary turbulence state where the forcing balances the dissipation. We use values of $T_c=500T_p$ and $C=\ln5/T_c$ (where $T_p$ is the peak period corresponding to $k_p$) which lead to favorable convergence rate in our study.

\section{Results}
\label{sec:numres}
In this section, we present the results from simulations of free-decay turbulence and forcing turbulence with broad and narrow bandwidths. The former is conducted at different effective steepness of the initial conditions, and the latter at different forcing amplitudes, in order to cover a sufficient range of nonlinearity levels. In the following, we will focus on the scaling of $S_\eta(k)$ with $k$ and $P$ at different nonlinearity levels, and investigate the mechanisms underlying the variations. 

\subsection{Spectral slopes}
\label{sec:spectra}
We first define the omnidirectional wavenumber spectrum $S_{\eta}(k)$ (we neglect its $t$ dependence for simplicity in the definition) by
\begin{equation}
    S_{\eta}(k)=\int_0^{2\pi} |\Tilde{\eta}(\bm{k})|^2 k d\theta,
    \label{eq:Sk}
\end{equation}
where $\bm{k}=(k_x,k_y)$ and $k=|\bm{k}|$, $\Tilde{\eta}(\bm{k})$ is the spatial Fourier transform of $\eta(\bm{x})$:
\begin{equation}
    \Tilde{\eta}(\bm{k})=\iint_{[0,2\pi]\times[0,2\pi]} \eta(\bm{x})e^{-i\bm{k}\cdot\bm{x}}d\bm{x}.
\end{equation}

To demonstrate the (quasi-) stationarity of the spectral evolution, we define two integral measures $E_{in}$ and $\Psi_{in}$ respectively for the spectra and compensated spectra (which assigns more weight on the high-wavenumber part), given by
\begin{equation}
    E_{in}(t)\equiv\int_{k_c}^{k_d} S_\eta(k,t)dk,
    \label{eq:E}
\end{equation}
\begin{equation}
    \Psi_{in}(t)\equiv\int_{k_c}^{k_d} k^{5/2}S_\eta(k,t)dk,
    \label{eq:psi}
\end{equation}
where $k_c=19$ locates within the inertial range as is shown later (also corresponding to the upper bound of the broadband forcing). We check the evolutions of $\Psi_{in}/E_{in}$ and $\Psi_{in}$ in free-decay and forcing cases to characterize their stationary state, where the denominator in the first quantity is used to account for the slow decay of energy level with time in the free-decay case.

\begin{figure}
  \centerline{\includegraphics[scale =0.36]{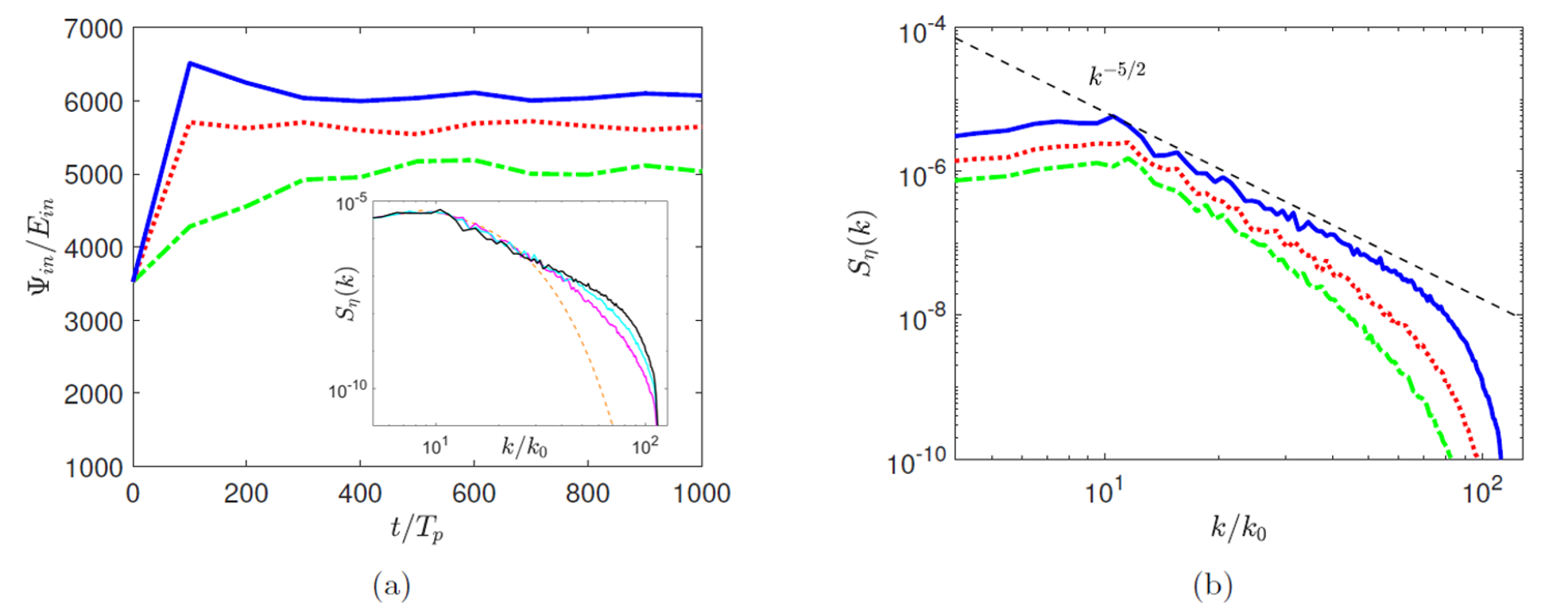}}% Images in 100% size
  \caption{(a) Time evolution of $\Psi_{in}/E_{in}$ and (b) stationary spectra in free-decay cases with $\epsilon=0.068$ ({\color{green}\dashdot}), $0.106$ ({\color{red}\dotL}) and $0.151$ ({\color{blue}\rule[0.5ex]{0.5cm}{0.25pt}}) evaluated at $t=500T_p$ from different runs. Inset of (a): wave spectra obtained at $t=0$ ({\color{orange}\dashL}), $100T_p$ ({\color{magenta}\rule[0.5ex]{0.5cm}{0.25pt}}), $200T_p$ ({\color{cyan}\rule[0.5ex]{0.5cm}{0.25pt}}) and $500T_p$ ({\color{black}\rule[0.5ex]{0.5cm}{0.25pt}}) with $\epsilon=0.151$. The theoretical power-law scaling $k^{-5/2}$ ({\color{black}\dashL}) is indicated in (b) for reference.}
\label{fig:Sfreedecay}
\end{figure}

The results obtained from free-decay cases are presented in figure \ref{fig:Sfreedecay}. Figure \ref{fig:Sfreedecay}(a) shows evolution of $\Psi_{in}/E_{in}$ with different nonlinearity levels, measured by the value of $\epsilon\equiv k_pH_s/2$ evaluated at $t=500T_p$. It can be seen that quasi-stationary states are established for all cases after $t=500T_p$. A typical spectral evolution for the case with $\epsilon=0.151$ is also shown as an inset to demonstrate the convergence of the spectrum to a power-law state. The spectra at quasi-stationary states for different nonlinearity levels (i.e., values of $\epsilon$) are shown in figure \ref{fig:Sfreedecay}(b). At high nonlinearity level of $\epsilon=0.151$, we observe a clear power-law spectrum which has slope $\alpha\approx -5/2$ consistent with WTT solution \eqref{WTT} in an approximate range of $[10,65]$. We note that the power-law range ends at $k\approx 65$ which is smaller than $k_d=115$, similar to other numerical simulations with a sharp dissipation cutoff \cite[e.g.][]{pan2014direct,dyachenko2004weak}, probably due to the interaction of the spectra with the dissipation range. With the decrease of nonlinearity level $\epsilon$, the power-law spectrum becomes shorter and steeper, reaching $\alpha\approx -3.4$ at $\epsilon=0.068$. This steepening of the spectra is in contrast to the results from Majda-McLaughlin-Tabak (MMT) turbulence with $\omega=k^2$ and quartet resonance \citep{hrabski2020effect}, mainly because the latter forms a continuous resonant system \citep{faou2016weakly} at low nonlinearity (we will elaborate this more in \S \ref{sec:tri}). We also remark that $\epsilon=0.151$ is about the highest nonlinearity we can reach in the current HOS context. Previous free-decay simulations \citep{onorato2002freely,yokoyama2004statistics} which result in $\alpha\approx -5/2$ are both conducted at a value of $\epsilon$ close to 0.15 (respectively 0.15 and 0.14).

\begin{figure}
  \centerline{\includegraphics[scale =0.55]{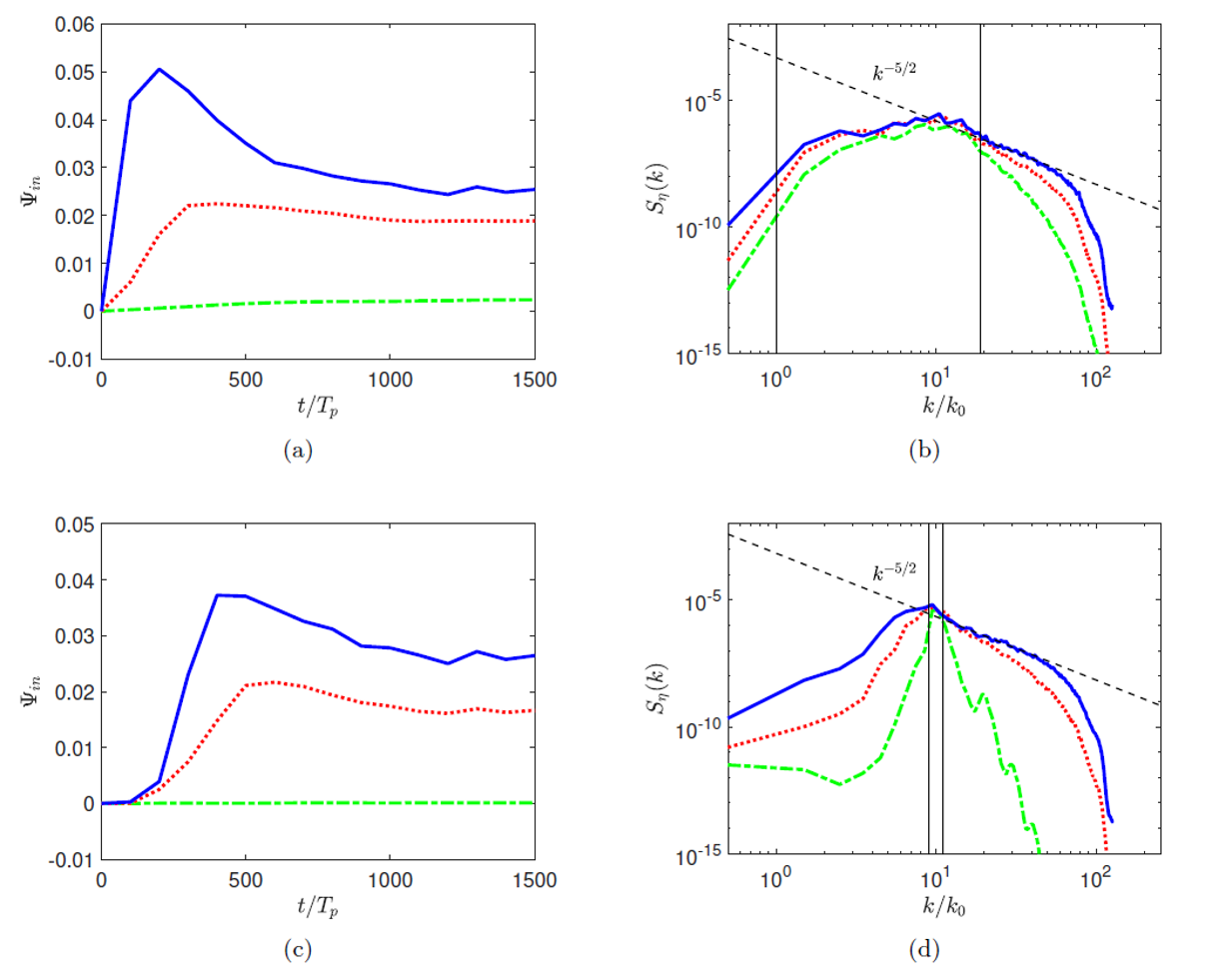}}% Images in 100% size
  \caption{(a) Time evolution of $\Psi_{in}$ and (b) stationary spectra in broadband forcing case with $f_0=3\times10^{-7}$, $\epsilon=0.061$ ({\color{green}\dashdot}); $f_0=1.6\times10^{-6}$, $\epsilon=0.118$ ({\color{red}\dotL}); $f_0=3.2\times10^{-6}$, $\epsilon=0.145$ ({\color{blue}\rule[0.5ex]{0.5cm}{0.25pt}}). (c) Time evolution of $\Psi_{in}$ and (d) stationary spectra in narrow-band forcing case with $f_0=8\times10^{-7}$, $\epsilon=0.059$ ({\color{green}\dashdot}); $f_0=4\times10^{-6}$, $\epsilon=0.114$ ({\color{red}\dotL}); $f_0=1\times10^{-5}$, $\epsilon=0.148$ ({\color{blue}\rule[0.5ex]{0.5cm}{0.25pt}}). The theoretical power-law scaling $k^{-5/2}$ ({\color{black}\dashL}) and boundaries of the forcing ranges ({\color{black}\rule[0.5ex]{0.5cm}{0.25pt}}) are indicated in (b) and (d) for reference.
  }
\label{fig:EnergySpectrum}
\end{figure}

The results from the forcing cases are shown in figure \ref{fig:EnergySpectrum}. The evolution of $\Psi_{in}$ in the narrow- and broadband cases with different forcing amplitudes $f_0$ (resulting in different values of $\epsilon$ at stationary state) are plotted in figure \ref{fig:EnergySpectrum}(a) and (c), all showing stationary states in $[1000T_p, 1500T_p]$. The stationary power-law spectra for respectively the two cases are plotted in (b) and (d). For both cases, we observe that the spectral slopes $\alpha$ approach $-5/2$ at sufficiently high forcing/nonlinearity (consistent with previous work \cite{dyachenko2004weak}). With the decrease of nonlinearity, the spectra for both cases become steeper, which are also reported in experiments in \cite{falcon2007observation,nazarenko_lukaschuk_mclelland_denissenko_2010,deike2015role}. However, we observe that the spectra at low nonlinearity clearly show a dependence on the bandwidth of the forcing, with the one from narrow-band case much steeper (even not showing a power law) than the one from broadband case. It can also be noticed that, for the narrow-band case, the spectrum at low nonlinearity level ($\epsilon=0.059$) exhibits discrete super-harmonic peaks, which will be explained through bound waves in \S \ref{sec:boundWaves}.  

\begin{figure}
  \centerline{\includegraphics[scale =0.6]{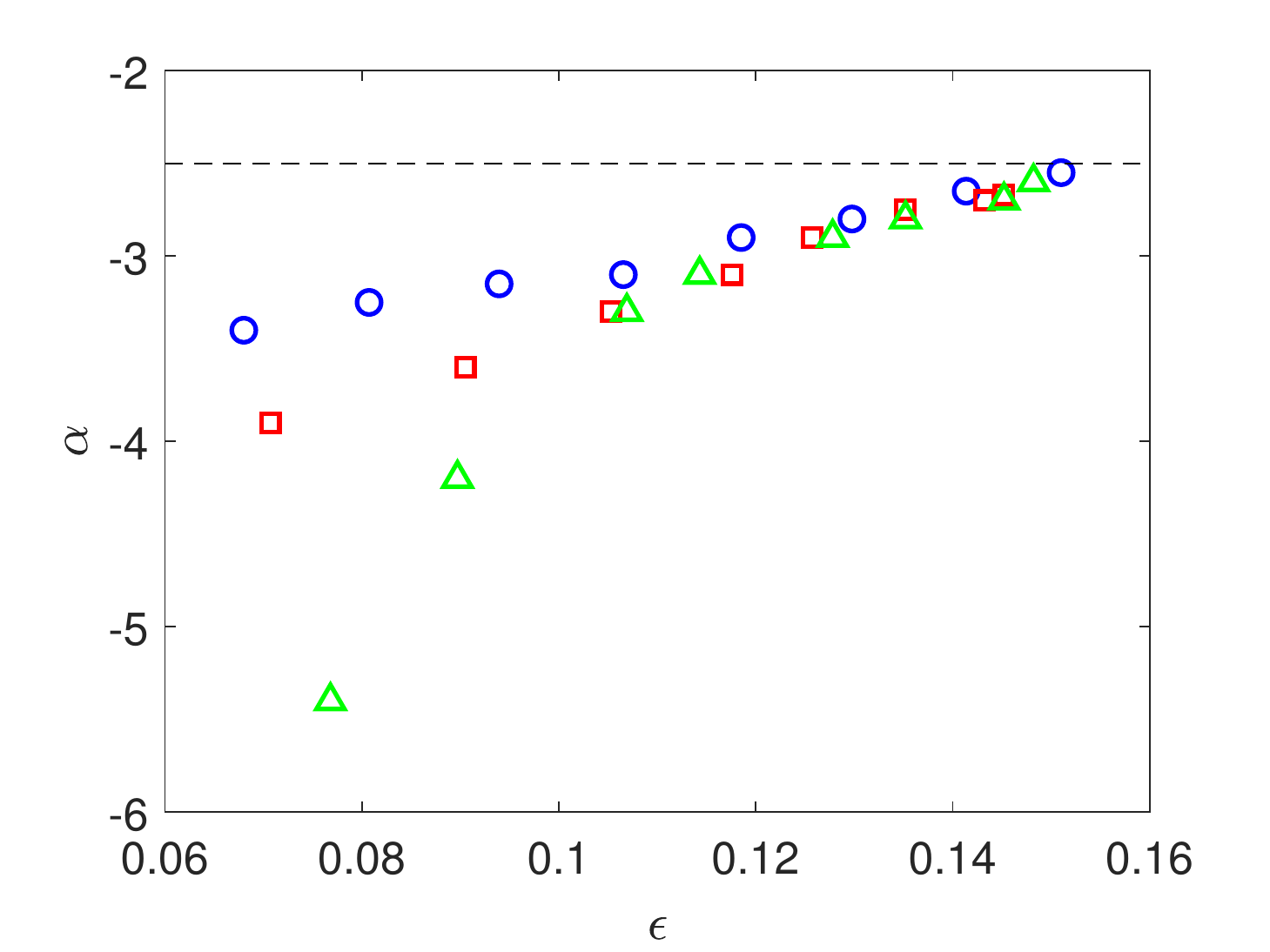}}% Images in 100% size
  \caption{The spectral slope $\alpha$ as functions of effective steepness $\epsilon$ for the free-decay ({\color{blue}$\circ$}), broadband forcing ({\color{red}$\Box$}) and narrow-band forcing ({\color{green}$\triangle$}) cases, compared to the prediction of WTT with $\alpha=-5/2$ ({\color{black}\dashL}).}
\label{fig:epsilonToAlpha}
\end{figure}

To obtain a complete view of spectral slopes for both the free-decay and forcing cases, we plot the values of $\alpha$ in all cases as functions of the nonlinearity level in figure \ref{fig:epsilonToAlpha}. This plot is limited above by the stability of HOS and below by the existence of a power-law spectra. In practice, we consider a power-law spectrum not existing if the power-law range is less than 0.5 decade or if the spectrum is dominated by discrete peaks such as the one with low nonlinearity level in figure \ref{fig:EnergySpectrum}(d). We see that for all cases, the spectral slope $\alpha$ approaches WTT value $-5/2$ at sufficiently high nonlinearity levels of $\epsilon\approx 0.15$. With the decrease of $\epsilon$, all spectra become steeper but with different steepening rates especially at relatively low nonlinearity level. It is also clear that the narrow-band forcing case shows a transition at $\epsilon_c \approx 0.11$ below which a very rapid steepening is observed. The mechanisms underlying these behaviors will be further analyzed in \S \ref{sec:boundWaves} and \S \ref{sec:tri}.

%We can also see that $\alpha$ is larger (less steeper spectrum) in the freely decaying cases than in the forcing cases at the same nonlinearity level, especially when the nonlinearity is weak. As for the less steeper spectrum for the freely decaying cases than the forcing cases, it may also be attributed to the effects of bound waves or the absence of the external force. The force may affect the energy cascade in the inertial range through nonlinear interactions with waves at small scales and lead to the deviation from WTT prediction.

\subsection{Energy flux}
\label{sec:flux}
In this section, we investigate the scaling between the spectral level of $S_\eta(k)$ and the energy flux $P$. For the evaluation of $P$, we can directly compute the energy transfer due to nonlinear terms in the primitive equation \citep{hrabski2020effect} or use energy dissipation rate as a measure of $P$ \cite[e.g.][]{pan2014direct}. For dissipation localized at high wavenumbers (such as our cases), the two approaches are equivalent (as all energy flux through the inertial range is dissipated at high wavenumbers in a stationary state). Therefore, we use dissipation rate as a measure of $P$, which takes the form \citep{pushkarev1996turbulence,pan2014direct}:
\begin{equation}
    P=\iint_{\bm{k}} \gamma_k(|\tilde{\eta}({\bm{k})}|^2+k|\widetilde{\phi^S}({\bm{k}})|^2)d\bm{k},
    \label{eq:flux}
\end{equation}
where $\widetilde{\phi^S}(\bm{k})$ is the spatial Fourier transform of $\phi^S(\bm{x})$. We note that \eqref{eq:flux} is slightly less accurate in the free-decay case since the spectra slowly evolves in the quasi-stationary state. However, the evolution rate of the spectra is much smaller than the energy flux so that the associated error is negligible.

%In addition, we define an integral measure on the spectral level by
%\begin{equation}
%    \tilde{E}\equiv\int_{k_c}^{k_d} S_\eta(k)dk,
%\end{equation}
%where $k_c=19k_0$ locates within the inertial range (also corresponding to the upper bound of the broadband forcing). 

\begin{figure}
  \centerline{\includegraphics[scale =0.6]{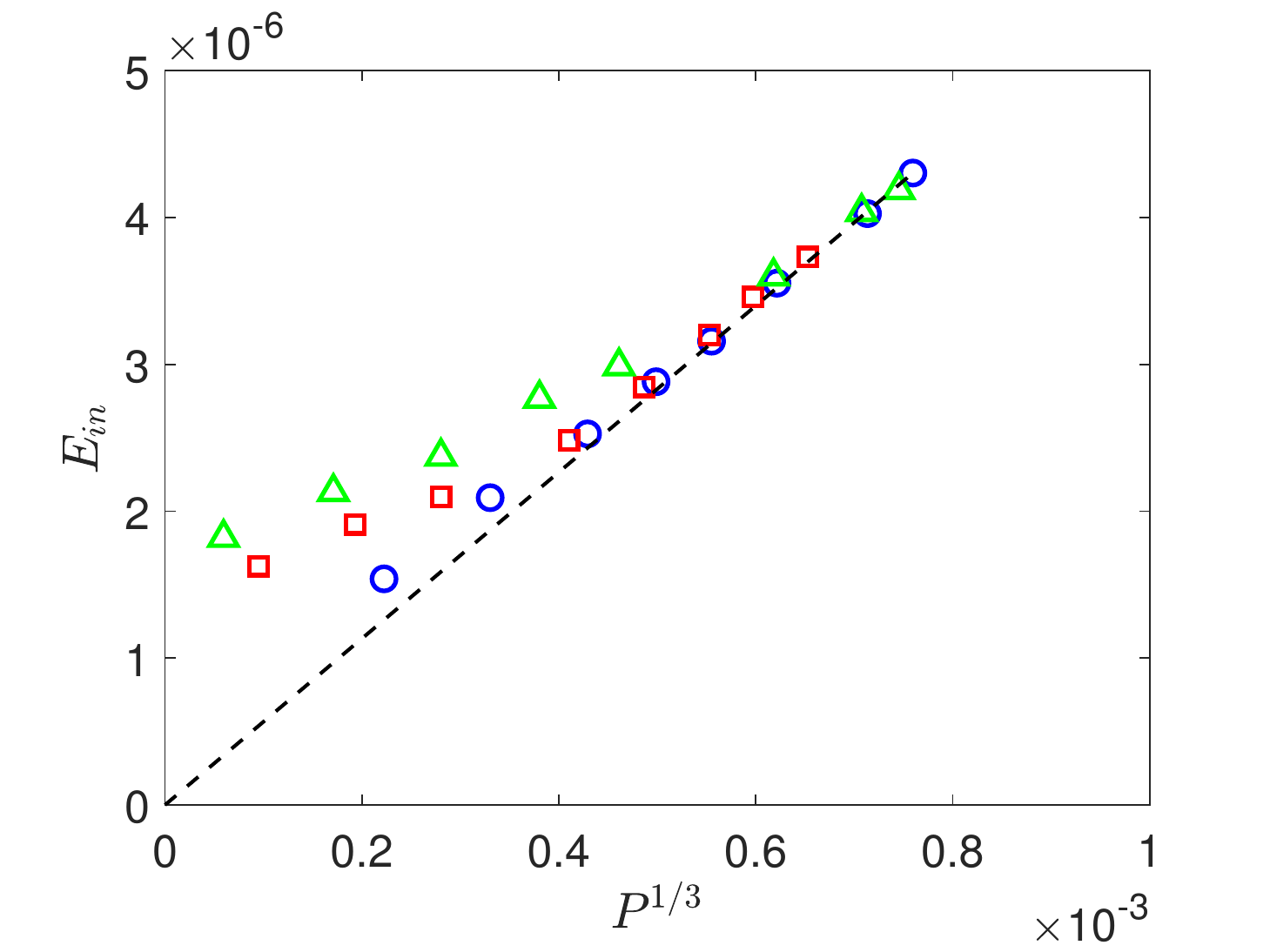}}% Images in 100% size
  \caption{Plots of the energy density in the inertial range $E_{in}$ as a function of $P^{1/3}$ for free-decay ({\color{blue}$\circ$}), broad-band forcing ({\color{red}$\Box$}) and narrow-band forcing ({\color{green}$\triangle$}) cases, comparing to the prediction of WTT $E_{in}\sim P^{1/3}$ ({\color{black}\dashL}).}
\label{fig:PE}
\end{figure}

The scaling between $E_{in}$ and $P^{1/3}$ is plotted in figure \ref{fig:PE} for the free-decay case and the two forcing cases. The variable $P^{1/3}$ is used for the horizontal axis so that WTT scaling $E_{in}\sim P^{1/3}$ becomes a straight line in the figure. We see that, at high nonlinearity level, the scaling in all cases approach the WTT scaling, although the range of consistency is longer in the free-decay and the broadband forcing cases. With the decrease of nonlinearity, the scaling deviates from the WTT prediction with a smaller value of $P$ for given $E_{in}$. This indicates a reduced capacity of energy cascade with the reduction of nonlinearity. As nonlinearity level is further decreased, all curves approach states with $P\rightarrow 0$ and finite $E_{in}$, suggesting the formation of ``frozen turbulence'', which is previously (only) introduced for capillary waves \citep{pushkarev2000turbulence,pan2014direct}. This result is remarkable because of the existence of exact resonances for gravity waves on a discrete grid of $\bm{k}$ (rational torus), in contrast to capillary waves. Before this study it is not clear whether the exact quartet resonances of gravity waves lead to a cascade as in MMT turbulence \citep{hrabski2020effect}. The results here suggest that the energy cascade should not be expected in spite of the existence of energy transfer within a small number of resonant quartets.

In the following, we will study hypothetical physical mechanisms about bound waves and finite-size effect, which may lead to the spectral behaviors presented in \S \ref{sec:spectra} and \S \ref{sec:flux}.

%\begin{figure}
%     \centering
%     \begin{subfigure}[b]{1\textwidth}
%         \centering
%         \includegraphics[trim=0cm 0cm 0cm 0cm, clip,scale=0.6]{PEBroad.eps}
%         \caption{}
%         \label{fig:PEBroad}
%     \end{subfigure}
%     %\hfill
%     \begin{subfigure}[b]{1\textwidth}
%         \centering
%         \includegraphics[trim=0cm 0cm 0cm 0cm, clip,scale=0.6]{PENarrow.eps}
%         \caption{}
%         \label{fig:PENarrow}
%     \end{subfigure}
%        \caption{Plots of the energy density in the inertial range $\tilde{E}$ as a function of $P^{1/3}$ ({\color{blue}\Lcirc}) for (a) broad-band forcing and (b) narrow-band forcing cases, comparing to the prediction of WTT $\tilde{E}\sim P^{1/3}$ ({\color{black}\dashL}).}
%        \label{fig:PE}
%\end{figure}

\subsection{Bound waves}
\label{sec:boundWaves}
In this section, we study the effect of bound waves, which is argued as a major factor influencing spectral behavior in previous experiments \citep{michel2018self}. Here we generally define bound waves as wave components that do not satisfy the linear dispersion relation, no matter if they are generated by carrier-wave distortion \citep{plant1999bound,plant2004bound}, super-harmonics \citep{herbert2010observation,cobelli2011different,michel2018self} or other mechanisms \citep{campagne2019identifying,longuet1992capillary} proposed in previous literature. In fact, as we will show in \S 3.3.1, all bound waves in a periodic-domain simulation can be interpreted through non-resonant nonlinear interactions. This definition distinguishes bound waves from free waves which satisfy the linear dispersion relation (or lie in its vicinity). To separate bound waves from the wave field, it is necessary to conduct a spatiotemporal analysis and obtain wavenumber-frequency spectra $S_\eta(k,\omega)$, defined as
\begin{equation}
    S_\eta(k,\omega)=\int_0^{2\pi} |\Tilde{\eta}(\bm{k},\omega)|^2 k d\theta,
    \label{eq:Swk}
\end{equation}
where $\Tilde{\eta}(\bm{k},\omega)$ is the spatiotemporal Fourier transform of $\eta(\bm{x},t)$:
\begin{equation}
    \Tilde{\eta}(\bm{k},\omega)=\iiint_{[0,T_L]\times[0,2\pi]\times[0,2\pi]} \eta(\bm{x},t)h_T(t)e^{-i(\bm{k}\cdot\bm{x}-\omega t)}d\bm{x}dt,
\label{eq:etaKW}
\end{equation}
with $h_T(t)$ the Tukey window \citep{bloomfield2004fourier} of length $T_L=20T_p$, the time duration of the collected data within $1480T_p\sim1500T_p$ at the stationary state.

\subsubsection{Generation mechanisms}
\begin{figure}
  \centerline{\includegraphics[scale =0.55]{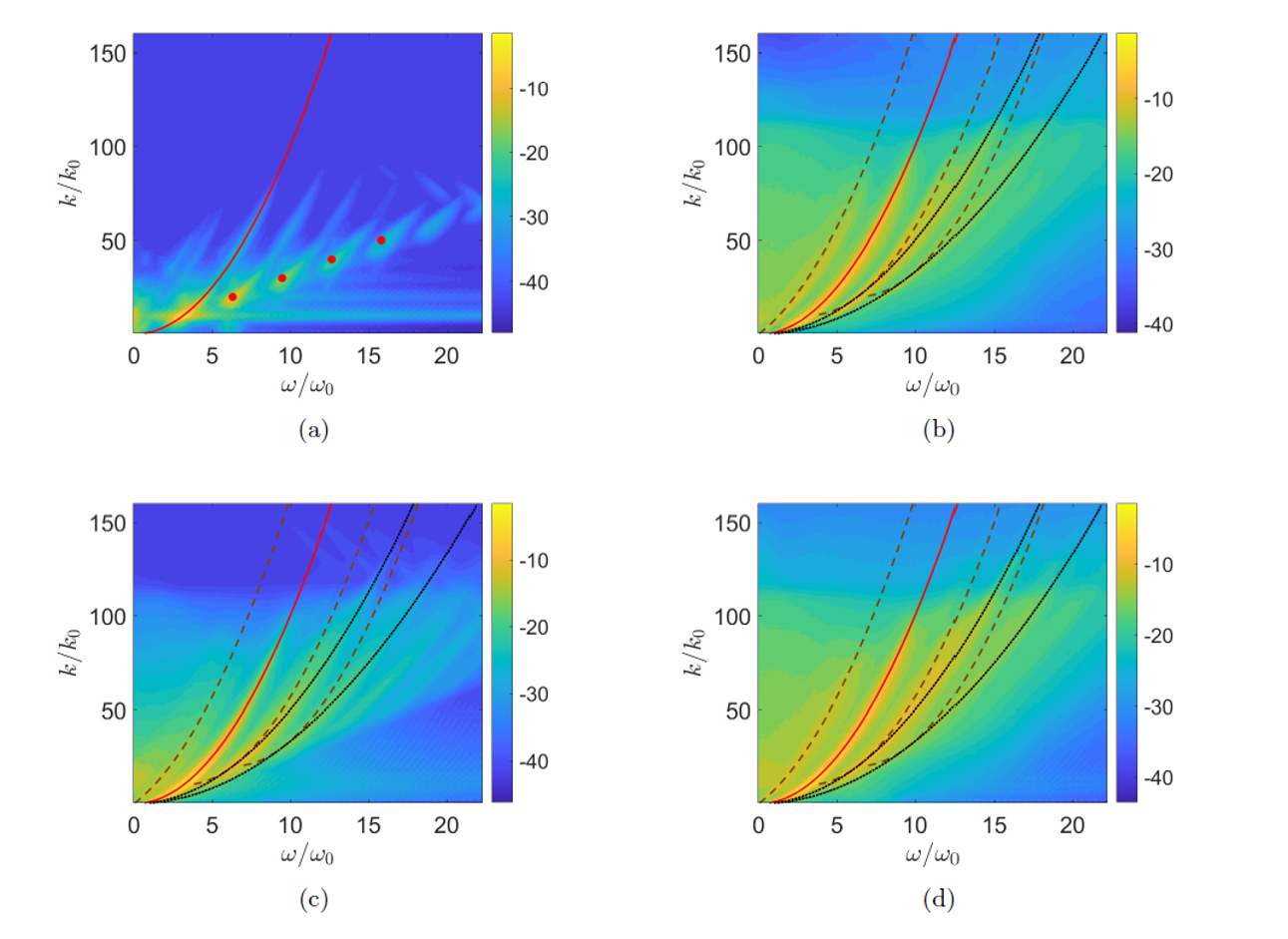}}% Images in 100% size
  \caption{Normalized wavenumber-frequency spectra $S_\eta(\omega,k)/S_\eta(\omega_p,k_p)$ in log scale for narrow-band forcing cases with (a) $\epsilon=0.059$, (b) $\epsilon=0.148$ and broadband forcing cases with (c) $\epsilon=0.061$, (d) $\epsilon=0.145$. The linear dispersion relation is marked by {\color{red}\rule[0.5ex]{0.5cm}{0.25pt}}. In (a), the peak modes are marked by {\color{red}$\bullet$} computed from \eqref{eq:b1} with $n= 2,3,4,5$. In (b-d), the lines corresponding to \eqref{eq:b2} with $m=2,3$ are indicated by {\color{black}\dotL}, and the lines corresponding to \eqref{eq:b3} with $l=-1,1,2$ are indicated by {\color{brown}\dashL}.}
\label{fig:Swk}
\end{figure}

The wavenumber-frequency spectra $S_\eta(k,\omega)$ are plotted in figure \ref{fig:Swk} from (a) to (d) for respectively narrow-band forcing with low and high nonlinearity levels, and broadband forcing with low and high nonlinearity levels. The results for free-decay cases are not shown since they are somewhat similar to the broadband forcing case. In all cases, we (seem to) observe significant amount of energy away from the linear dispersion relation (red curve in each sub-figure), indicating the presence of bound waves in the simulations. In addition, the plot for narrow-band forcing at low nonlinearity (figure \ref{fig:Swk}(a)) shows discrete peaks of bound wave components, a visually distinct pattern from other sub-figures. In retrospect to results in \S \ref{sec:spectra}, these discrete peaks correspond to the peaks in wavenumber spectra with low nonlinearity in figure \ref{fig:EnergySpectrum}(d). As marked by red dots in figure \ref{fig:Swk}(a), these peaks can be quantified as super-harmonics of the peak mode $(\omega_p, k_p)$ of carrier waves, in the form of
\begin{equation}
    (\omega_b,k_b)=(n\omega_p,nk_p),\quad n=2,3,4,...
    \label{eq:b1}
\end{equation}
The phase velocity for all bound wave components, described by \eqref{eq:b1}, can be computed as $\omega_b/k_b=\omega_p/k_p$, which is the same as the phase velocity of the peak mode of carrier waves. Therefore, the bound waves can be considered as non-dispersive, i.e., they are ``bound'' to the carrier wave as the latter travels. This behavior agrees with some of the classical views of bound waves described in \cite{lake1978new,plant2003new}.

For cases in \ref{fig:Swk}(b)-(d), however, the bound-wave patterns are dramatically different from that in (a). It is clear from the plots that these bound wave components are not necessarily non-dispersive, which is consistent with other general view of bound waves \cite[e.g.][]{phillips1981dispersion}. In these cases, we observe several bound-wave branches (in contrast to the main branch of linear dispersion relation) on the spectra $S_\eta(\omega,k)$. These bound-wave branches can be explained through two different mechanisms discussed below.

The first mechanism corresponds to the super-harmonics of all free waves, i.e., modes $(\omega,k)$ in the main (carrier) branch of the linear dispersion relation, satisfying 
\begin{equation}
    (\omega_b,k_b)=(m\omega,mk),\quad m=2,3,4,...
    \label{eq:b2}
\end{equation}
The second mechanism corresponds to bound waves generated by the (non-resonant) interactions between the harmonics of peak mode $(\omega_p,k_p)$ and an arbitrary free wave $(\omega,k)$ in the main branch, satisfying 
\begin{equation}
    (\omega_b,k_b)=(\omega+l\omega_p,k+lk_p),\quad l=\pm1,\pm2,\pm3 ...
    \label{eq:b3}
\end{equation}

The curves described by \eqref{eq:b2} and \eqref{eq:b3} are marked in \ref{fig:Swk}(b)-(d) respectively by black solid lines and brown dashed lines. The two mechanisms co-explain the super-branches of bound waves on the right of the main branch, as they both contribute to energy in each of the visible super-branches. The second mechanism with $l=-1$ further explains the sub-branches of bound waves on the left of the main branch. We remark that the two mechanisms \eqref{eq:b2} and \eqref{eq:b3} are also separately observed in experiments of \cite{herbert2010observation,cobelli2011different} and \cite{campagne2019identifying}. Our results here provide a more comprehensive view of bound waves: we need to consider the property of carrier waves to distinguish cases in \ref{fig:Swk}(a) and (b)-(d), and consider combined mechanisms \eqref{eq:b2} and \eqref{eq:b3} for explanation in the latter case.

%\begin{figure}
%     \centering
%     \begin{subfigure}[b]{1\textwidth}
%         \centering
%         \includegraphics[trim=0cm 0cm 0cm 0cm, clip,scale=0.6]{SwkBoundFree.eps}
%         \caption{}
%         \label{fig:Swk}
%     \end{subfigure}
%     %\hfill
%     \begin{subfigure}[b]{1\textwidth}
%         \centering
%         \includegraphics[trim=0cm 0cm 0cm 0cm, clip,scale=0.6]{SwkProportion.eps}
%         \caption{}
%         \label{fig:Proportion}
%     \end{subfigure}
%        \caption{(a) Typical normalized wavenumber-frequency spectrum $S_\eta(k,\omega)/S_\eta(k_p,\omega_p)$ plotted in log scale for a broad-band forcing case with $\epsilon=0.069$. The red solid curve represents the linear dispersion relation. The black dotted lines indicate bound waves generated as the harmonics of the free waves. The white dash-dot lines indicate bound waves generated from the nonlinear interaction between free waves and harmonics of the spectral peak. (b) The free-wave band on the wavenumber-frequency spectrum in the vicinity of linear dispersion relation. The red dashed lines are the left and right boundaries of the band $[\omega_k-\delta\omega,\omega_k+\delta\omega]$. The region bounded by black dash-dot lines corresponds to the inertial ranges for frequency and wavenumber spectrum used to calculate the proportion of bound wave energy to the total energy.}
%        \label{fig:BoundFree}
%\end{figure}

\subsubsection{Effects on wave spectra}
\begin{figure}
  \centerline{\includegraphics[scale =0.6]{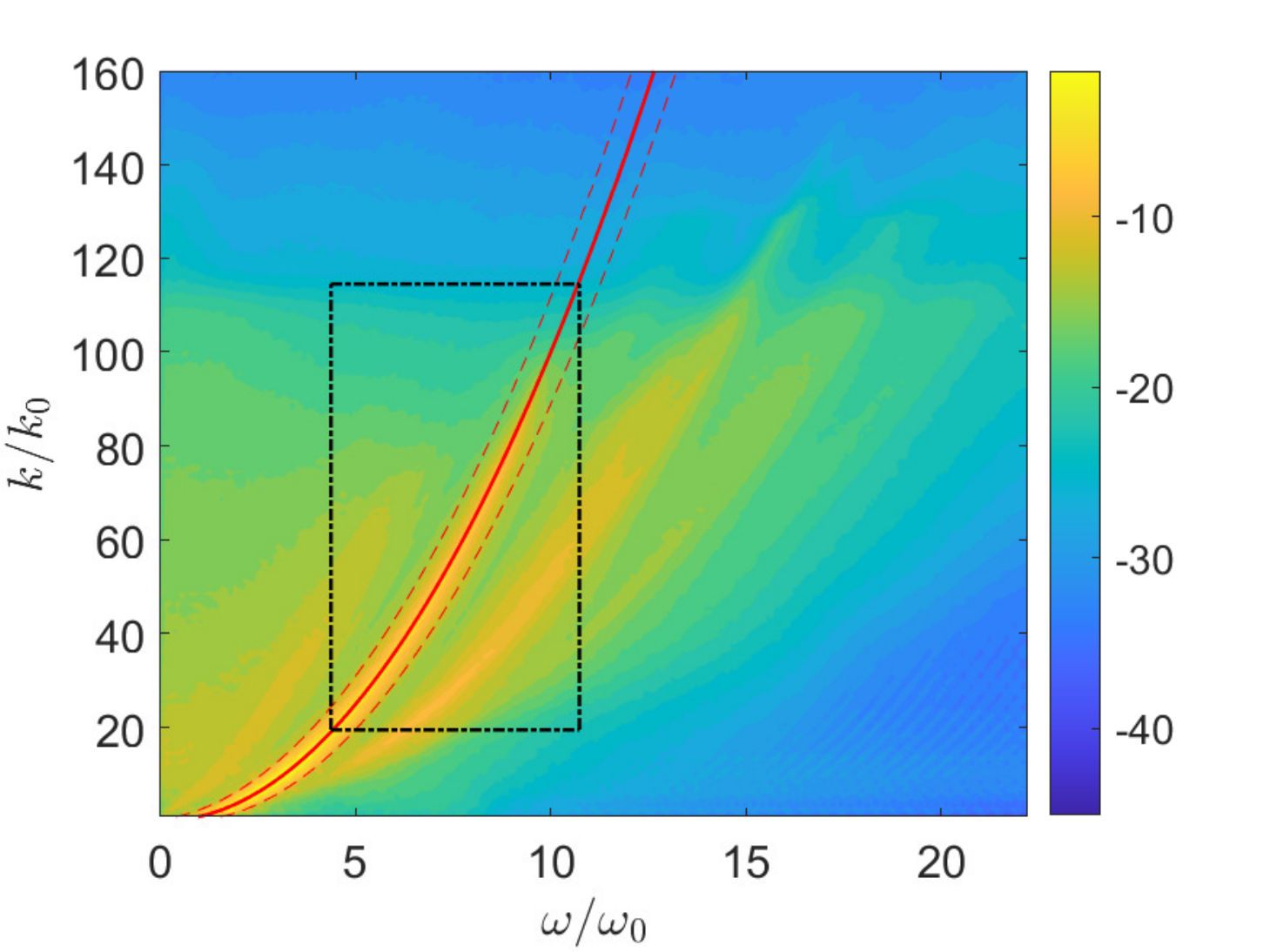}}% Images in 100% size
  \caption{A typical illustration for the separation of free and bound waves in a wavenumber-frequency spectrum. The region of $[\omega_k-\delta\omega,\omega_k+\delta\omega]$ defined in \eqref{eq:filter} is marked by {\color{red}\dashL}. The region bounded by {\color{black}\dashdot} corresponds to the inertial ranges  $[\omega_c,\omega_d]\times[k_c,k_d]$ used in \eqref{eq:Eb}.}
\label{fig:SwkProportion}
\end{figure}

Our next goal is to separate bound waves and free waves to elucidate their relative importance to the wave turbulence of gravity waves. The algorithm for the separation is illustrated in figure \ref{fig:SwkProportion}, which shows the definition of a free-wave finite band (by dashed line) in the vicinity of the linear dispersion relation. Specifically, this finite band is generated through a filter (similar to \cite{campagne2019identifying})
\begin{equation}
    f(\bm{k},\omega)\equiv\left\{
    \begin{array}{lc}
    1,& |\omega-\omega_k| \leq \delta \omega \\
    0,& |\omega-\omega_k| > \delta \omega
    \end{array}
    \right.
\label{eq:filter}
\end{equation}
where $\delta \omega=0.6\omega_0$ is a parameter characterizing the width of the free-wave band, with $\omega_0=1$ the fundamental frequency in the domain. We have tested that this choice of $\delta \omega$ lies in a stationary range in terms of the results discussed below, i.e., reducing or increasing it results in respectively insufficient free-wave energy or contamination by bound-wave branches. For the separation, we apply the filter \eqref{eq:filter} directly to $\Tilde{\eta}(\bm{k},\omega)$ and obtain the free-wave and bound-wave components as $\Tilde{\eta}_f(\bm{k},\omega)=f(\bm{k},\omega)\Tilde{\eta}(\bm{k},\omega)$ and $\Tilde{\eta}_b(\bm{k},\omega)=\Tilde{\eta}(\bm{k},\omega)-\Tilde{\eta}_f(\bm{k},\omega)$. Then we compute the free-wave spectra $S_{\eta}^f(k,\omega)$, $S_{\eta}^f(k)$ and bound-wave spectra $S_{\eta}^b(k,\omega)$, $S_{\eta}^b(k)$ likewise using \eqref{eq:Swk} and \eqref{eq:Sk}. We note that it is also possible to directly apply the filter \eqref{eq:filter} to $S_\eta(k,\omega)$ for the separation, but our operation (w.r.t $\Tilde{\eta}$) is desirable due to the study that will be discussed in \S \ref{sec:tri}.

\begin{figure}
  \centerline{\includegraphics[scale =0.67]{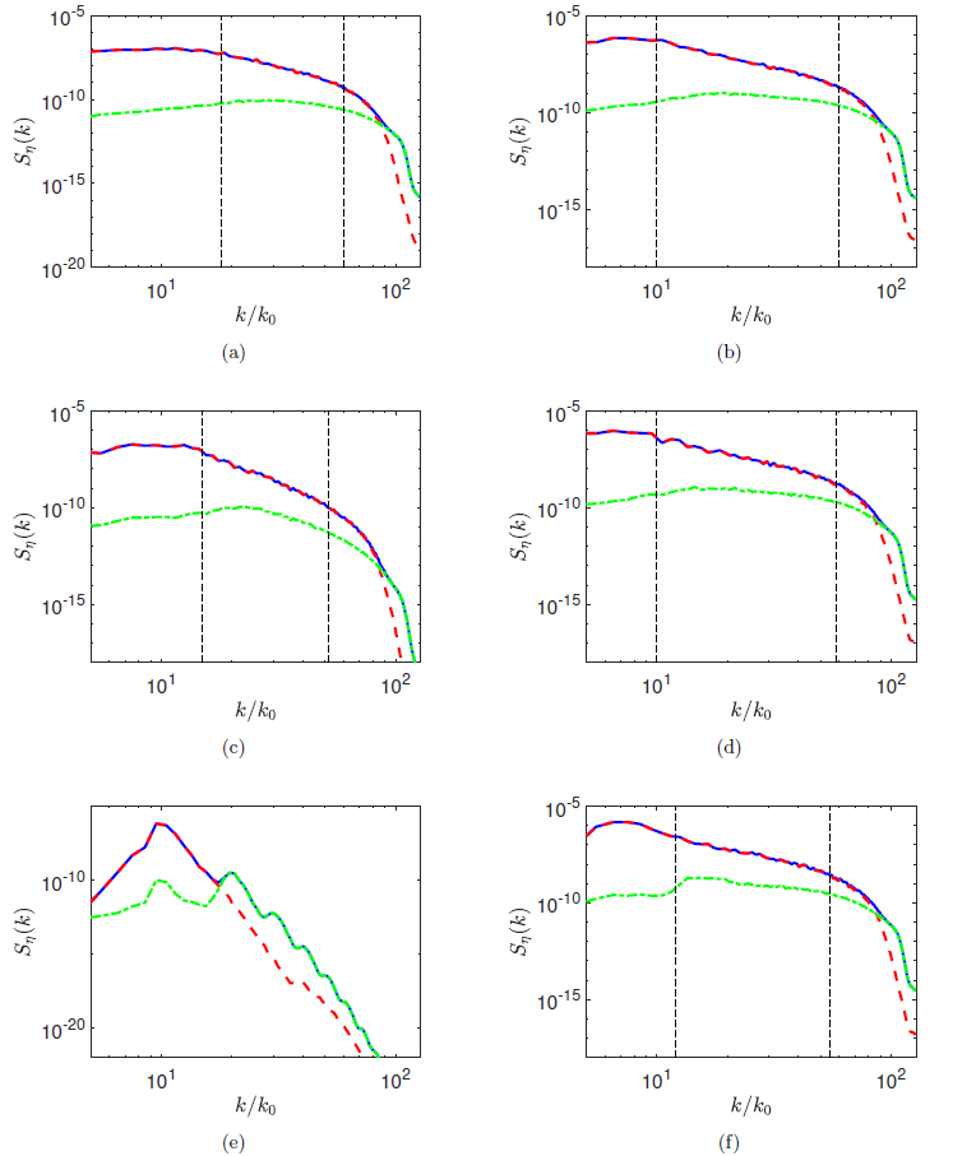}}% Images in 100% size
  \caption{$S_\eta^f(k)$ ({\color{red}\dashL}), $S_\eta^b(k)$ ({\color{green}\dashdot}) and $S_\eta(k)$ ({\color{blue}\rule[0.5ex]{0.5cm}{0.25pt}}) in free-decay cases with (a) $\epsilon=0.068$ and (b) $\epsilon=0.151$; broadband forcing cases with (c) $\epsilon=0.071$ and (d) $\epsilon=0.145$; narrow-band forcing cases with (e) $\epsilon=0.059$ and (f) $\epsilon=0.148$. The boundaries of the power-law ranges are indicated by {\color{black}\dashL}, except for (e) where discrete peaks are observed.}
\label{fig:SkSeparate}
\end{figure}

The free-wave and bound-wave wavenumber spectra are shown in figure \ref{fig:SkSeparate} for all three (free-decay, braodband and narrow-band forcing) cases at high and low nonlinearity levels. In most cases, we find that the power-law range of the spectra are dominated by free-wave components with their energy at least one order of magnitude higher than those of the bound-wave components. The only exception is the case with narrow-band forcing at low nonlinearity level shown in figure \ref{fig:SkSeparate}(e), where bound waves dominate most of the range above $k\approx 17$ exhibiting discrete peaks.

The general behavior of bound waves in figure \ref{fig:SkSeparate} suggest that they are not the major factor causing steepening of the spectra except in the narrow-band forcing case. This point will be made more clear after further quantifying the fraction of bound waves in the wave field. For this purpose, we define the bound-wave energy in the inertial range as
\begin{equation}
    E_b\equiv\int_{\omega_c}^{\omega_d}\int_{k_c}^{k_d} S_{\eta}^b(k,\omega)dkd\omega
    \label{eq:Eb}
\end{equation}
where $[k_c, k_d]$ and $[\omega_c=k_c^{1/2}, \omega_d=k_d^{1/2}]$ are a box describing the limits of inertial range as shown in figure \ref{fig:SwkProportion}. We quantify the fraction of bound waves using the ratio of $E_b/E_t$, where $E_t$ is the total wave energy in the inertial range computed in a similar way as \eqref{eq:Eb} but with $S_{\eta}(k,\omega)$ replacing $S_{\eta}^b(k,\omega)$.

\begin{figure}
  \centerline{\includegraphics[scale =0.6]{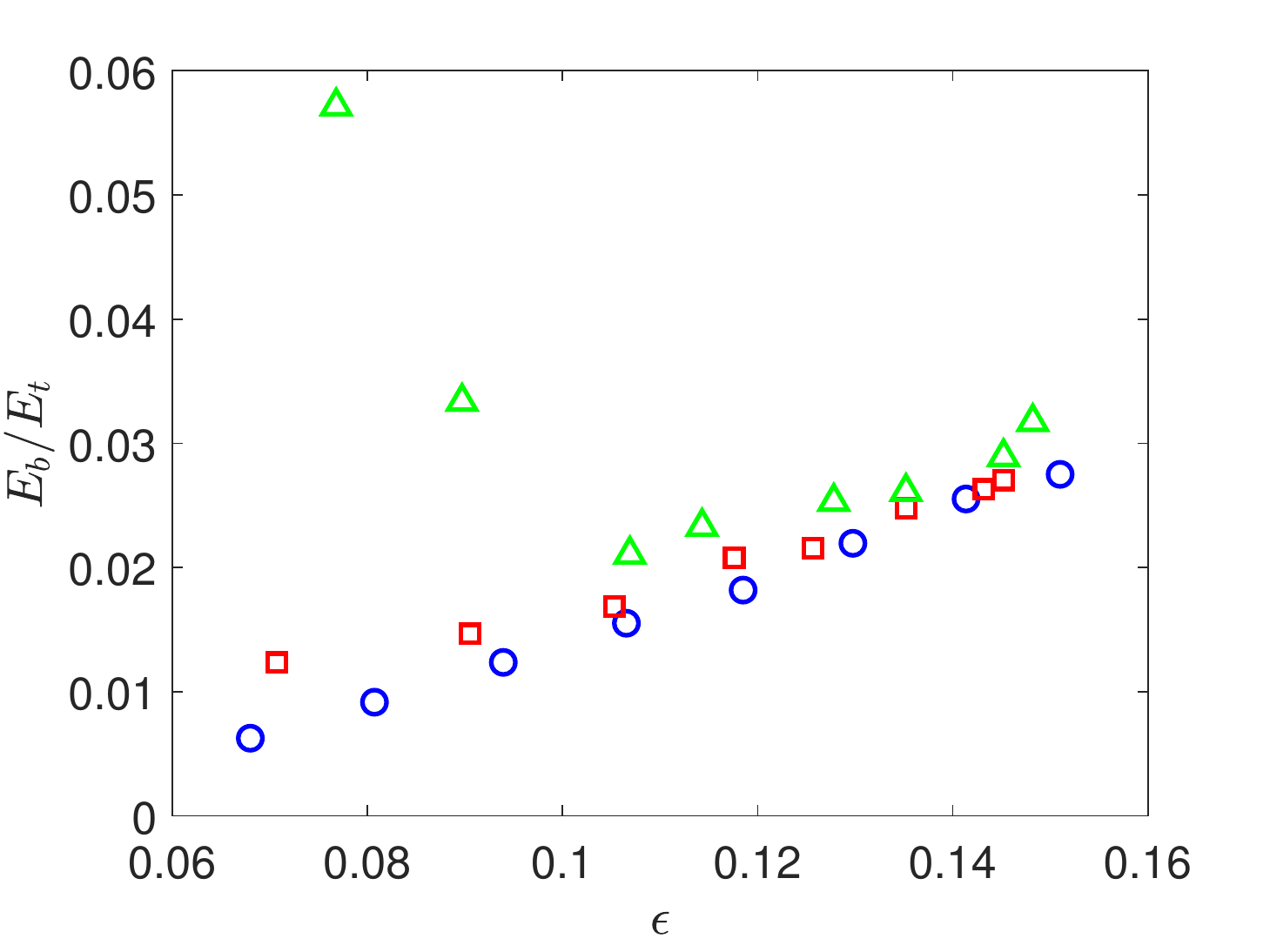}}% Images in 100% size
  \caption{The proportions of bound wave energy $E_b/E_t$ as functions of $\epsilon$ for the free-decay ({\color{blue}$\circ$}), broadband forcing ({\color{red}$\Box$}) and narrow-band forcing ({\color{green}$\triangle$}) cases.}
\label{fig:rb}
\end{figure}

Figure \ref{fig:rb} summarizes the quantity $E_b/E_t$ at different nonlinearity levels for all three cases. For the broadband forcing and free-decay cases, we see that the fraction of bound-wave energy consistently decreases with the decrease of nonlinearity level. This provides a clear evidence that the steepening of the spectra with decrease of nonlinearity is not caused by bound waves in these cases. For narrow-band forcing case, the fraction of bound-wave energy is consistently larger than those in the other two cases (at all nonlinearity levels). In addition, with the decrease of nonlinearity, a transition occurs at a critical nonlinearity level ($\epsilon_c\approx0.11$) below which $E_b/E_t$ increases with the decrease of $\epsilon$. (In general, we expect that the critical level $\epsilon_c$ to increase with the decrease of the forcing bandwidth.) The critical level $\epsilon_c$ is consistent with the transition in figure \ref{fig:epsilonToAlpha} below which a much more rapid steepening of the spectral slope is observed for the narrow-band forcing case. We thus conclude that the presence of bound waves explains the largest steepening rate of the spectra at low nonlinearity levels in the narrow-band forcing case. Moreover, the significant fraction of bound waves in the narrow-band forcing case also provide an explanation of the rapid deviation from $E_{in}\sim P^{1/3}$ shown in figure \ref{fig:PE}, since bound waves account for energy which does not follow the WTT cascade pathway.

\subsection{Finite-size effect}
\label{sec:tri}
In order to further understand the steepening of spectra and reduced energy flux capacity at low nonlinearity levels, we investigate another hypothetical mechanism of finite-size effect. In general, finite-size effect arises due to the violation of the assumption of the infinite domain in WTT \citep{pushkarev2000turbulence,lvov2006discreteness,nazarenko2006sandpile}. In the framework of WTT kinetic equation, the energy cascade is enabled by modes on the continuous resonant manifold satisfying (for gravity waves)
\begin{equation}
    \bm{k}_1+\bm{k}_2=\bm{k}_3+\bm{k},
\label{eq:resonantk}
\end{equation}
\begin{equation}
    \omega_1+\omega_2=\omega_3+\omega,
\label{eq:resonantw}
\end{equation}
where $\omega_i^2=k_i\ (i=1,2,3)$. In a finite domain, discreteness of wavenumber and frequency (imposed by the domain boundary) reduces the manifold defined by \eqref{eq:resonantk} and \eqref{eq:resonantw} by limiting the resonances to discrete points. This discreteness effect can be compensated by the nonlinear broadening which allows the occurrence of quasi-resonances, characterized by a modification of \eqref{eq:resonantw} as
\begin{equation}
    |\omega_1+\omega_2-\omega_3-\omega|<\Delta \omega,
\end{equation}
with $\Delta \omega$ a nonlinear broadening parameter depending on the nonlinearity level.

One way to quantify the finite-size effect (in particular to measure the nonlinear broadening and interaction strength) of gravity waves is through a tricoherence analysis \cite[see e.g.][for bi-coherence study for capillary waves]{pan_yue2017understanding}. We first define a tricoherence function 
\begin{equation}
    T(\bm{k},\bm{k}_1,\bm{k}_2)\equiv\frac{|\langle\Tilde{\eta}_f(\bm{k}_1,t)\Tilde{\eta}_f(\bm{k}_2,t)\Tilde{\eta}_f^*(\bm{k}_3,t)\Tilde{\eta}_f^*(\bm{k},t)\rangle|}{\langle|\Tilde{\eta}_f(\bm{k}_1,t)||\Tilde{\eta}_f(\bm{k}_2,t)||\Tilde{\eta}_f^*(\bm{k}_3,t)||\Tilde{\eta}_f^*(\bm{k},t)|\rangle}
\label{eq:TkDef}
\end{equation}
where $\bm{k}_3 \equiv -\bm{k}+\bm{k}_1+\bm{k}_2$, $\langle\cdot\rangle$ denotes the time average (for stationary states), and $*$ denotes the complex conjugate. $\Tilde{\eta}_f(\bm{k},t)$ corresponds to the free-wave modes, and is computed by the spatial Fourier transform of $\eta_f(\bm{x},t)$ with the latter obtained after applying the filter \eqref{eq:filter} to the wavenumber-frequency spectra. By definition, the function $T(\bm{k},\bm{k}_1,\bm{k}_2)$ continuously varies between 0 and 1, with 1 corresponding to exact resonance among $\bm{k}$, $\bm{k}_1$, $\bm{k}_2$ and $\bm{k}_3$. In \eqref{eq:TkDef}, we use $\Tilde{\eta}_f$ instead of $\Tilde{\eta}$ for the evaluation to remove the contamination by bound waves since they result in noisy values of $T$ not lying in the vicinity of the resonant manifold (as found in previous work \cite{pan_yue2017understanding}). This operation is also compatible with findings in \S \ref{sec:boundWaves} that bound waves do not contribute to the general steepening of the spectra (except in the narrow-band forcing case).

\begin{figure}
  \centerline{\includegraphics[scale =0.37]{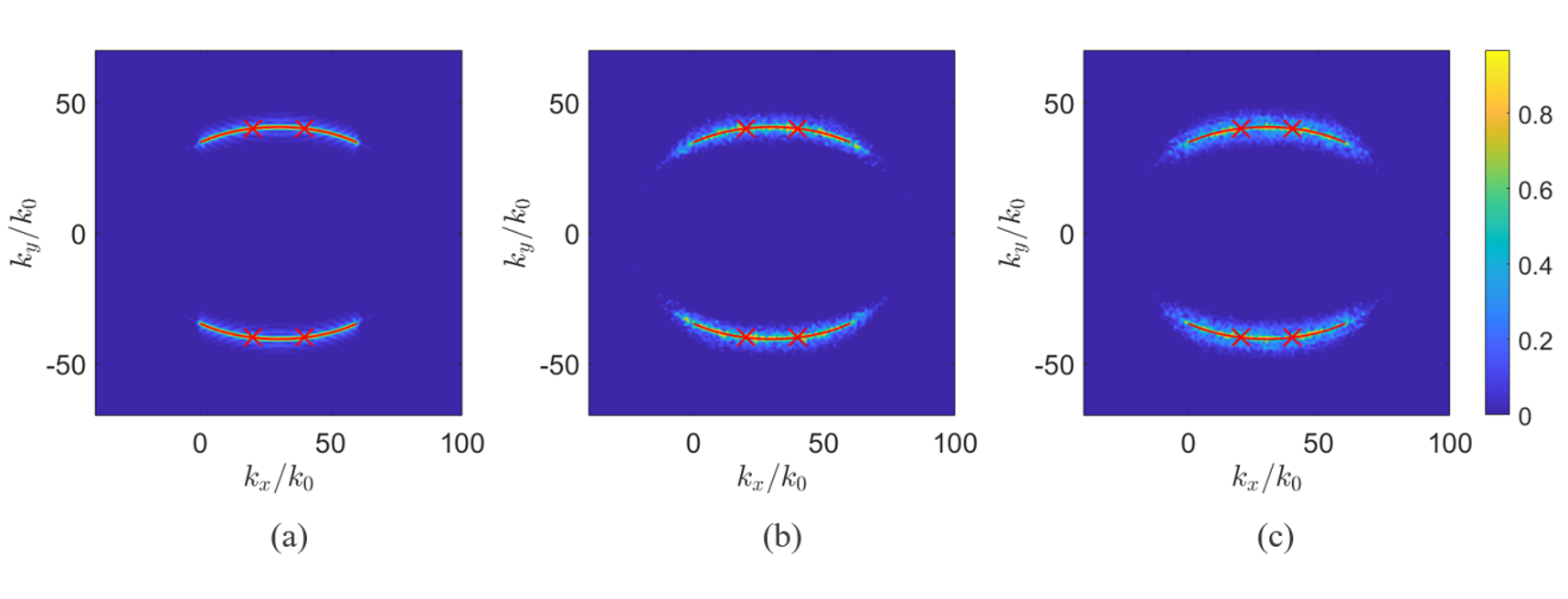}}% Images in 100% size
  \caption{The tricoherence $T(\bm{k},\bm{k}_1,\bm{k}_2)$ with $\bm{k}_1=(40,40)$, $\bm{k}_2=(20,-40)$ for broad-band forcing cases with (a) $\epsilon=0.071$, (b) $\epsilon=0.118$ and (c) $\epsilon=0.145$. The continuous resonant manifold is plotted by {\color{red}\rule[0.5ex]{0.5cm}{0.25pt}}. The discrete resonant solutions with $\bm{k}\in(\mathbb{Z},\mathbb{Z})$ are marked by {\color{red}$\times$}. The data for generating these figures are collected from $t=1460T_p$ to $t=1500T_p$ in the stationary state.}
\label{fig:Tk}
\end{figure}

To facilitate the visualization of $T$, we fix $\bm{k}_1=(40,40)$ and $\bm{k}_2=(20,-40)$ so that $T(\bm{k})$ with $\bm{k}=(k_x,k_y)$ can be shown by two-dimensional contour plots. Figure \ref{fig:Tk} shows such plots of $T(\bm{k})$ in the broadband forcing case at three different nonlinearity levels. The results for the narrow-band forcing and free-decay cases are not included due to their similarity to the presented results (the quantification of broadening and interaction strength for all cases will be presented later). Also shown in figure \ref{fig:Tk} is the continuous resonant manifold (red lines) satisfying \eqref{eq:resonantk} and \eqref{eq:resonantw}, and discrete resonant points (red crosses) with $\bm{k}$ only taking integer values. In particular, to compute the resonant manifold we need to consider the wave traveling direction and complex conjugate relation for real functions, with details presented in Appendix A.

From figure \ref{fig:Tk} we can see that all significant values of $T$ are concentrated close to the resonant manifold (as a result of using $\Tilde{\eta}_f$ in \eqref{eq:TkDef}).  It is also clear that the nonlinear broadening is visibly wider at higher nonlinearity level compared to that at lower nonlinearity level. Furthermore, the discrete resonant points seem to be sparse on the continuous manifold, with all four points in each sub-figure of figure \ref{fig:Tk} as trivial quartet solutions (either with repetition or symmetry with given vectors of $\bm{k}_1$ and $\bm{k}_2$). This sparsity can be further demonstrated by a numerical analysis which identifies no non-trivial solution with stretching of the current resonant manifold (see Appendix B). This problem has also been considered analytically and numerically in \cite{kartashova2006fast,lvov2006discreteness}, which reach a consistent conclusion on the rareness of the discrete solutions except for two special types of collinear quartets and tridents (which however do not exist in our case with waves propagating only to the positive $x$ direction). Moreover, the sparsity of discrete solutions in this case is in contrast to the situation of $\omega=k^2$ MMT dispersion relation studied in \cite{hrabski2020effect}. The difference in the structure of the discrete resonant solutions is the key for the gravity-wave spectra and MMT spectra to show completely different behaviors at low nonlinearity, with the former deviating from WTT spectral slope but the latter approaching WTT spectral slope.

%From figure \ref{fig:Tk} we can see that all significant values of $T$ are concentrated close to the resonant manifold (as a result of using $\Tilde{\eta}_f$ in \eqref{eq:TkDef}).  It is also clear that the nonlinear broadening is visibly wider at higher nonlinearity level compared to that at lower nonlinearity level. Furthermore, the discrete resonant points seem to be sparse on the continuous manifold, which is in contrast to the situation of $\omega=k^2$ MMT dispersion relation studied in \cite{hrabski2020effect}. This difference is the key for the two systems to show completely different behaviors at low nonlinearity, with gravity-wave spectrum deviate from WTT solution but MMT spectrum approaching WTT solution.

To further quantify the effect of nonlinearity level to quartet resonances, we define two measures $\hat{L}_b$ and $I$ respectively for the broadening width and overall interaction strength. For $\hat{L}_b$, we define it as a characteristic width \citep{pan_yue2017understanding}:
\begin{equation}
    \hat{L}_b\equiv\frac{\sum\limits_{\bm{k}} |\hat{\Omega}|T(\bm{k})}{\sum\limits_{\bm{k}} T(\bm{k})},
\label{eq:Lb}
\end{equation}
where $|\hat{\Omega}|$ is the normalized frequency mismatch given by $|\hat{\Omega}|=|\Omega|/k^{-1/2}$ with $\Omega\equiv\omega_1+\omega_2-\omega_3-\omega_k$ and the denominator $k^{-1/2}$ estimating the frequency discreteness at $k$ associated with the wavenumber spacing. In equation \eqref{eq:Lb}, the summation is for all the grid points of $\bm{k}$. Thus, $\hat{L}_b$ measures the nonlinear broadening around the exact solutions by the first moment of $T(\bm{k})$.

We further define $I$ as
\begin{equation}
    I\equiv\sum\limits_{\bm{k}} T(\bm{k}),
    \label{eq:I}
\end{equation}
which measures the overall interaction strength by summing up contributions from both resonant and quasi-resonant interactions.

\begin{figure}
  \centerline{\includegraphics[scale =0.4]{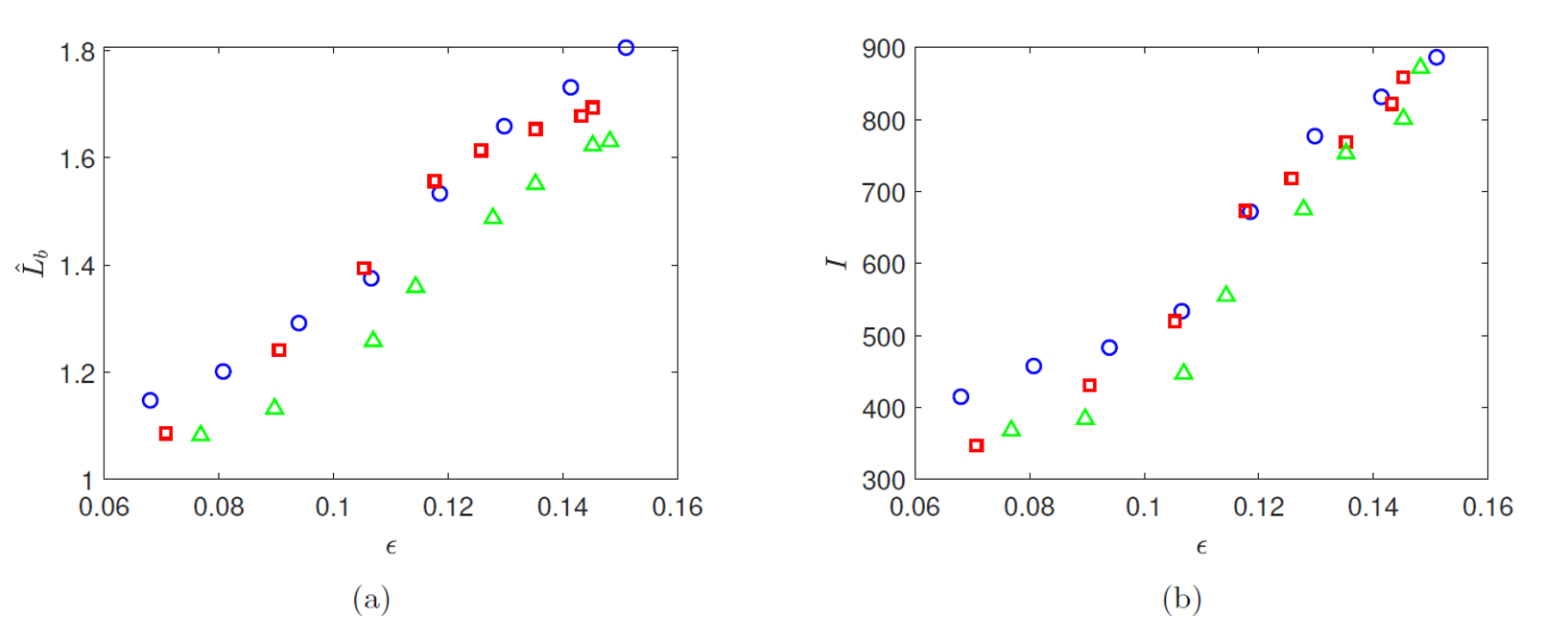}}% Images in 100% size
  \caption{(a) The characteristic length of nonlinear broadening $\hat{L}_b$ and (b) the intensity of nonlinear interactions $I$ as functions of $\epsilon$ for the free-decay ({\color{blue}$\circ$}), broadband forcing ({\color{red}$\Box$}) and narrow-band forcing ({\color{green}$\triangle$}) cases.}
\label{fig:LbI}
\end{figure}

The characteristic width $\hat{L}_b$ and the interaction strength $I$ are plotted as functions of $\epsilon$ for all cases in figure \ref{fig:LbI}. We can observe a general increase of $\hat{L}_b$ and $I$ for increasing $\epsilon$, indicating a wider nonlinear broadening and a larger interaction strength with more quasi-resonances alleviating the finite-size effect. We remark that this clear behavior of $\hat{L}_b$ and $I$ are only possible to resolve by using $\tilde{\eta}_f$ to evaluate the tri-coherence. The reduction of nonlinear broadening and interaction strength with decreasing nonlinearity is consistent with steepening of spectra and reduction of energy flux capacity discussed in \S \ref{sec:spectra} and \S \ref{sec:flux}. Therefore, we conclude that the finite-size effect is a major contributor to the spectral behaviors at low nonlinearity level (especially for cases with sufficient spectral bandwidth). Finally, the values of $\hat{L}_b$ and $I$ in the narrow-band forcing case are slightly smaller than the other two cases for all nonlinearity levels, probably because of the higher fraction of bound waves (which leads to a larger $\epsilon$ than the other two cases, but not larger $\hat{L}_b$ or $I$).

%The characteristic length $\hat{L}_b$ is plotted as a function of $\epsilon$ for all cases. We can observe a general increase of $\hat{L}_b$ for increasing $\epsilon$, indicating a wider nonlinear broadening with more quasi-resonances alleviating the finite-size effect. The reduction of nonlinear broadening with decreasing nonlinearity is consistent with steepening of spectra and reduction of energy flux capacity discussed in \S \ref{sec:spectra} and \S \ref{sec:flux}. Therefore, we conclude that the finite-size effect is a major contributor to the spectral behaviors at low nonlinearity level (especially for cases with sufficient spectral bandwidth). Finally, the values of $\hat{L}_b$ in the narrow-band forcing case are slightly smaller than the other two cases for all nonlinearity levels, probably because of the higher fraction of bound waves (which leads to a larger $\epsilon$ than the other two cases, but not larger $\hat{L}_b$).

\section{Conclusions}
\label{sec:conc}
We conduct numerical simulation of Euler equations to study the surface gravity wave turbulence in three representative conditions, namely free-decay, narrow-band forcing and broadband forcing turbulence. In all cases, We find that the scaling of the wave spectra with wavenumber and energy flux both approach the WTT solution at sufficiently high nonlinearity level. With the decrease of nonlinearity level, steeper spectra and reduced energy flux capacity can be observed indicating deviation from WTT, with the largest deviation rate found in the narrow-band forcing case. Two hypothetical mechanisms on bound waves and finite-size effect to explain these spectral variations are investigated. For bound waves, we elucidate their generation mechanisms through a spatiotemporal analysis (which generalizes the previous study on this topic) and find that their fraction generally decreases with the decrease of nonlinearity level except for the narrow-band forcing case. This suggests that bound waves only account for the rapid deviation from WTT in the narrow-band forcing case (but not for the other two cases). For finite-size effect, we perform a tri-coherence analysis and find that both the nonlinear broadening and interaction strength decrease with the decrease of nonlinearity level, which accounts for the deviation from WTT at low nonlinearity level in all cases of our simulation. We finally remark that cautions have to be taken in applying these numerical findings to experiments due to the additional complexity inevitably involved in the latter.

\section*{Declaration of Interests}
The authors report no conflict of interest.

\appendix
\section{Computation of resonant manifold in figure \ref{fig:Tk}}\label{appA}
We start by considering a (quasi-)stationary wave field described by 
\begin{equation}
    \eta(\bm{x},t) = \sum_i A_i e^{i(\bm{\kappa}_i \cdot \bm{x} -\omega_i t )} + A_i^* e^{-i(\bm{\kappa}_i \cdot \bm{x} -\omega_i t )},
    \label{eq:field}
\end{equation}
where $A_i=|A_i|e^{i\phi_i} \in \mathbb{C}$ and $\omega_i \in \mathbb{R}^+$. Since the wave fields in our simulations generally only contain waves traveling to the positive $x$ direction (due to the setting of initial condition and forcing), for each mode in \eqref{eq:field} we have $\kappa_{ix}\cdot\omega_i>0$ so that $\bm{\kappa}_i=(\kappa_{ix},\kappa_{iy})\in (\mathbb{R}^+,\mathbb{R})$.

Considering $\Tilde{\eta}(\bm{k},t)$ with $\bm{k}\in(\mathbb{R},\mathbb{R})$ as the spatial Fourier transform of $\eta(\bm{x},t)$, we can see that
\begin{equation}
    \Tilde{\eta}(\bm{k}_i,t)=\left\{
    \begin{array}{lc}
    A_i e^{-i \omega_i t }, & k_{ix} > 0 \\
    A_i^* e^{i\omega_i t }, & k_{ix} < 0
    \end{array},
    \right.
    \label{eq:etai}
\end{equation}
Substituting \eqref{eq:etai} to \eqref{eq:TkDef}, we obtain 
\begin{equation}
T=|\langle e^{i(-\Omega_d t + \Phi_d)} \rangle|,
\label{eq:TOmega}
\end{equation}
where $\Phi_d=\phi_1+\phi_2-\phi_3-\phi$, and
\begin{equation}
    \Omega_d=\sign(k_{1x}) \omega_1+\sign(k_{2x})\omega_2-\sign(k_{3x})\omega_3-\sign(k_x)\omega,
    \label{eq:omegad}
\end{equation}
with
\begin{equation}
    \sign(x)=\left\{
    \begin{array}{lc}
    1, & x > 0 \\
    -1, & x < 0
    \end{array},
    \right.
    \label{eq:sign}
\end{equation}

\begin{figure}
  \centerline{\includegraphics[scale =0.5]{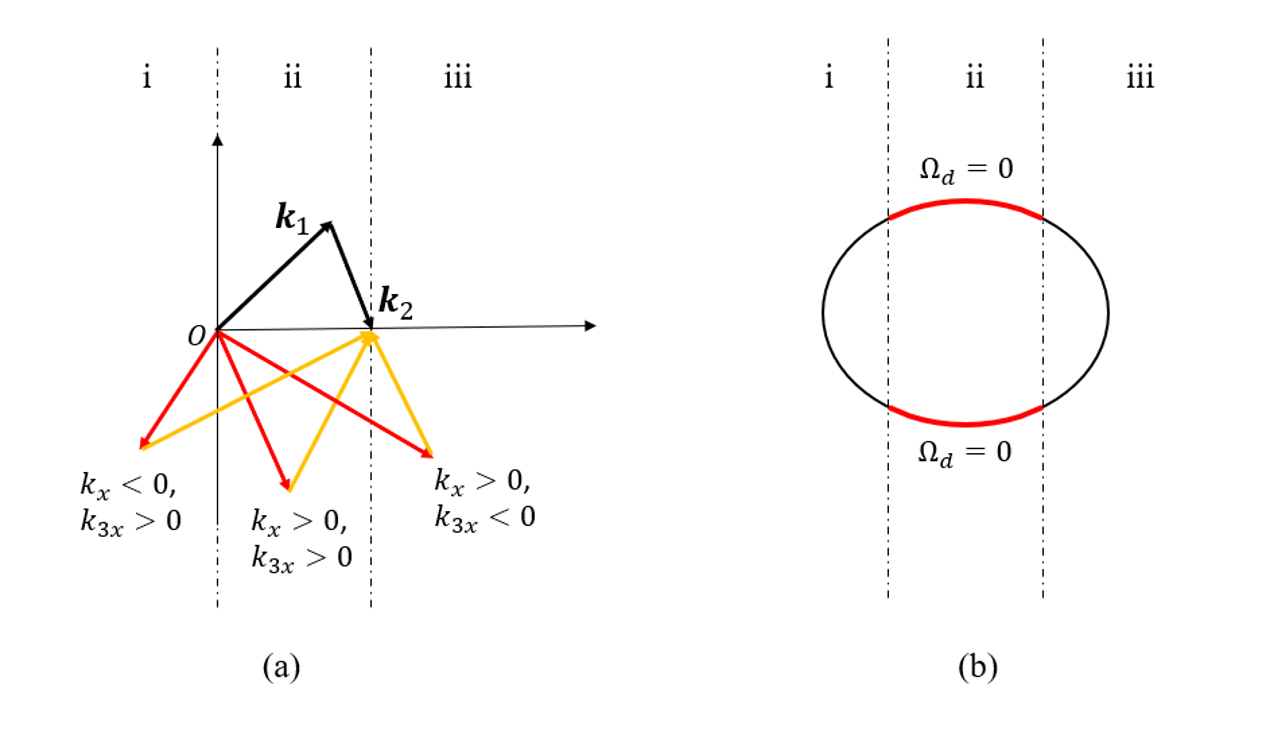}}% Images in 100% size
  \caption{(a) Sketches of the three situation in discussion. Vectors $\bm{k}_1$ and $\bm{k}_2$ are indicated in the figure, $\bm{k}$ and $\bm{k}_3$ are denoted by red and yellow arrows. The two vertical lines marked the boundaries of situation (i), (ii) and (iii) in the discussion. (b) The solution in situation (ii) as denoted by red lines in the figure (ellipse bounded by the two vertical line).}
\label{fig:TkRes}
\end{figure}

In our case, we have $k_{1x}=40>0$, $k_{2x}=20>0$ and $k_{3x}=k_{1x}+k_{2x}-k_x=60-k_x$, so that \eqref{eq:omegad} is reduced to $\Omega_d=\omega_1+\omega_2-\sign(60-k_x)\omega_3-\sign(k_x)\omega$. We are interested in the case of $\Omega_d=0$, which results in $T=1$ indicating the resonant manifold. Whether $\Omega_d=0$ can be realized needs to be discussed in the following three situations (summarized in figure \ref{fig:TkRes}(a)):
\begin{enumerate}
    \item For $k_x<0$, we need $\Omega_d=\omega_1+\omega_2-\omega_3+\omega=0$ which has no solution (for $\omega\in \mathbb{R}^+$) on the $(k_x,k_y)$ plane.
    \item For $0<k_x<60$, we need $\Omega_d=\omega_1+\omega_2-\omega_3-\omega=0$ which has solutions shown as an ellipse bounded by $k_x=0$ and $k_x=60$ in figure \ref{fig:TkRes}(b). This solutions corresponds to the resonant manifold shown in figure \ref{fig:Tk}.
    \item For $k_x>60$, we have $\Omega_d=\omega_1+\omega_2+\omega_3-\omega=0$ which has no solution on the $(k_x,k_y)$ plane.
\end{enumerate}

In summary, the resonant manifold shown in figure \ref{fig:Tk} corresponds to the second case above, which is the only possibility to have solutions in $\Omega_d=0$ for a wave field traveling to the positive $x$ direction.

\section{Sparsity of the discrete resonant solutions}\label{appB}
We count the number of discrete resonant solutions as in figure \ref{fig:Tk}, but with stretched $\bm{k}_1$ and $\bm{k}_2$. The purpose is to demonstrate the sparsity of discrete resonances with the expansion of the resonant manifold. In particular, we define $\bm{k}_1=\lambda(40,40)$ and $\bm{k}_2=\lambda(20,-40)$ where $\lambda$ is a factor of stretching, and compute the number of exact resonances $N$ as a function of $\lambda$. For $\lambda$ varying from 1 to 10, our numerical searching algorithm gives $N=4$ always, indicating that no non-trivial solution can be identified. While this numerical study is presented for $\bm{k}_1$ and $\bm{k}_2$ in particular directions, we have also tested other directions to confirm that the exact resonances are indeed sparse for gravity waves.

\bibliographystyle{jfm}
\bibliography{main}

\end{document}